\documentclass[aip,apl,preprint]{revtex4-1}
\usepackage{graphicx}
\usepackage{color}
\usepackage{dcolumn}
\usepackage{bm}
\usepackage{amsmath, amssymb, stmaryrd}

\newcommand{\alb}[1]{\textcolor{red}{#1}}

\definecolor{linkcolor}{rgb}{0,0,0.6} 

\begin{document}
\small

\title{ \large New method of determination of the Fr\'eedericksz threshold \\ based upon precise fluctuation measurement} 



\author{A. Caussarieu}
\email[]{aude.caussarieu@ens-lyon.fr}
\affiliation{
Universit\'e de Lyon \\
Ecole Normale Sup\'erieure de Lyon, Laboratoire de Physique ,\\
C.N.R.S. UMR5672,  \\ 46, All\'ee d'Italie, 69364 Lyon Cedex
07,  France\\}
\author{A. Petrosyan}
\affiliation{
Universit\'e de Lyon \\
Ecole Normale Sup\'erieure de Lyon, Laboratoire de Physique ,\\
C.N.R.S. UMR5672,  \\ 46, All\'ee d'Italie, 69364 Lyon Cedex
07,  France\\}\affiliation{
Universit\'e de Lyon \\
Ecole Normale Sup\'erieure de Lyon, Laboratoire de Physique ,\\
C.N.R.S. UMR5672,  \\ 46, All\'ee d'Italie, 69364 Lyon Cedex
07,  France\\}
\author{S. Ciliberto}
\email[]{sergio.ciliberto@ens-lyon.fr}
\affiliation{
Universit\'e de Lyon \\
Ecole Normale Sup\'erieure de Lyon, Laboratoire de Physique ,\\
C.N.R.S. UMR5672,  \\ 46, All\'ee d'Italie, 69364 Lyon Cedex
07,  France\\}

\date{\today}

\begin{abstract}

In this paper we report a new method for determining  the critical threshold of the Fr\'eedericksz transition driven by an electric field. It is based on the measurement of the amplitude of the molecule fluctuations as a function of the voltage difference applied to a planar nematic cell. The precise measurement of the director fluctuations  of the liquid crystal is made possible by the use of  a very precise and sensitive  polarization interferometer.  The great advantage of the method is that it does not depend on complex fits as it is usually done in literature. 

\end{abstract}

\pacs{???}

\maketitle 

The Fr\'eedericksz Transition (FT) is widely used in modern optoelectronic devices based on liquid crystal technology such as amplitude phase modulators  and displays. For the development of such equipments, one needs precise knowledge of the different constants of the chosen liquid crystal in interaction with a specific substrate. The determination of these constants is generally achieved with several techniques that involve at one point the knowledge of the critical threshold of the FT \cite{Bradshaw, faetti-bogi}.
In this paper we propose a new method for  determining this last quantity : it is  based on  a very sensitive interferometer which allows a precise measurement of the  thermal fluctuation amplitude of the LC  molecule orientation. This   method does not need the knowledge of any  other parameters such as the thickness of the cell or the anchoring characteristics. 

\begin{figure}[htbp]
 \includegraphics[width=.5\columnwidth]{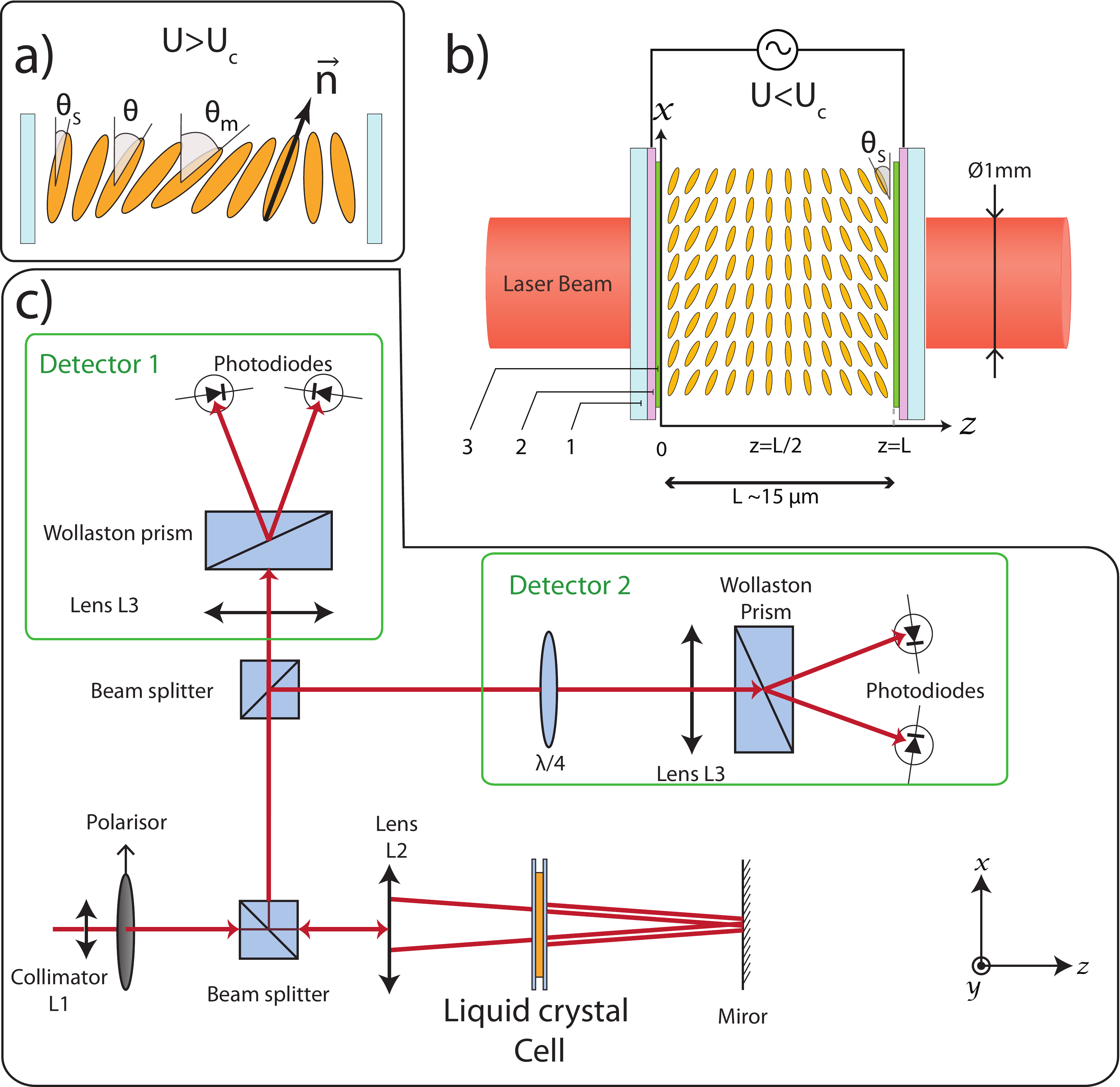}%
 \caption{Experimental set-up. a) Definition of the nematic director $\vec{n}$ ; b)Scheme of the liquid crystal cell used for the experiments : 1 - glass plate 2 - ITO 3 - rubbed PVA layer;  c)  Scheme of the quadrature phase interferometer}%
 \label{fig1}
 \end{figure} 

First, let us recall the main properties of the FT, a transition that occurs when a confined nematic liquid crystal (NLC) is submitted to an external electric, magnetic or optical field\cite{DeGennes, Oswald}. The order parameter of the FT is the unit pseudo vector ~$\vec{n}$ (the director) which defines the local direction of alignment of the molecules as represented on fig\ref{fig1}a). 
In the case of the FT driven by an electric field, the system under consideration is confined between two glass plates, separated by a distance $L$ (about $15\mu$m). The surfaces in contact with LC molecules are successively coated by ITO, for applying an electric field, and by a polymer layer which is rubbed along a specific direction to align the molecules in the vicinity of the plates.{ In our case, a planar anchoring\cite{cognard} is realized and the molecules are in average in the (x,z) plane, making a $\theta_s$ angle with the plates called pre-tilt angle (see fig\ref{fig1}a ). An anti-parallel configuration\cite{yeh-book} is chosen ($\theta(x=0)=\theta_s=-\theta(x=L)$) in which, in the absence of any external field, the system is totally symmetric (see fig\ref{fig1}b).}
 Applying a voltage difference $U$ between the electrodes, the NLC is submitted to an electric field $\vec{E}$ perpendicular to the plates ; to avoid polarization, the applied voltage is modulated at a frequency $f$ of a few kHz, $U(t)=U_0\sqrt{2}\cos2\pi f t$. When $U_0$ exceeds a critical value $U_c$, the planar state becomes unstable and the molecules rotate to align with the electrical field -- when the dielectric anisotropy of the LC is positive.  This transition is usually described as a second order phase transition between two orientational orders\cite{Oswald,DeGennes}. 
In our experiment the liquid crystal is the 5CB (4'-pentyl-4-biph\'enylcarbonitrile) whose parameters  are know with a good accuracy  (see for example ref.\cite{faetti-bogi} ). The polymer films are made of PVA (PolyVinyl Alcohol), which insures a strong anchoring on the plates of the 5CB molecules, {with $\theta_s \simeq 0.05rad$ and an anchoring energy  $W=3 \ 10^{-4} J/m^2$.}

\vspace{0.3cm}

In order to understand our measuring  technique one has to recall how the FT is theoretically described in the framework of the Franck-Oosen continuum theory of liquid crystals \cite{franck}. In such a description the order parameter of the transition is the angle $\theta$ between $\vec{n}$ and the initial uniform orientation of the molecules $\vec{x}$. 
In our configuration (planar cell), assuming strong anchoring ({\sl e.g.} that the director is fixed on the surfaces), the free energy per unit surface $F_s$ can be written as an integral over the thickness $L$ of the cell as shown in equation (\ref{eq:energie libre}). 

\begin{equation}
F_s=\frac{1}{2}\int_0^L\left\{(k_1(1+\kappa)\sin^2\theta)\theta_z^2-\vec{D}\cdot\vec{E}\right\}dz
\label{eq:energie libre}
\end{equation}

This expression takes into account both the electric ($\vec{E}$ is the electric field and $\vec{D}$ is the electric displacement vector) and elastic contributions \cite{Deuling} ; $k_1$ is an elastic constant and $\kappa$ characterizes the elastic anisotropy.  

The equilibrium description of the FT is given by the relation between the order parameter at the middle of the cell $\theta_m$ and the voltage $U_0$. This relationship  of $\theta_m$ as a function of $U_0$ can be obtained following two different methods. 
The first one\cite{Deuling} consists in computing a numerical  minimization (Num-min) of the free energy eq (\ref{eq:energie libre}) . The result of Num-min depends on the free parameter $U_c$ and this technique is often used to get $U_c$ from the experimental data \cite{Bradshaw,Deuling,faetti-bogi} 
The second method to determine analytically $U_c$ is to work in the vicinity of the threshold and to assume the sinusoidal form of the solution before expanding the free energy to find the equilibrium condition. This second method, in addition to the measurement of the critical threshold,  allows one to show the second order phase transition character of the FT, as will be shown on the following.
{In fact, assuming perfect boundary conditions (e.g. $\theta_s=0$ and infinite anchoring energy) and the sinusoidal form of the solution, $\theta=\theta_m(t)\sin(\frac{\pi z}{L})$, the series development of the free energy of the system (\ref{eq:energie libre}) becomes }: 

\begin{equation}
F_s=\frac{\pi^2k_1}{2L}\left[-\frac{\epsilon_\bot U^2}{\pi^2k_1}-\frac{\theta_m^2}{2}\varepsilon+\frac{\theta_m^4}{8}\left(\kappa+1+\varUpsilon\right)\right]
\label{eq:dvp_serie_mathematica}
\end{equation}

{In this expression the dielectric permittivities are defined along ($\epsilon_\sslash $) or perpendicular ($\epsilon_\bot$) to the molecular axis. 
The dielectric anisotropy is characterized by the difference $\epsilon_a=\epsilon_\sslash-\epsilon_\bot$ or by the non-dimensional constant $\varUpsilon=\epsilon_a/\epsilon_\bot$. }
The control parameter $\varepsilon$ is defined as $\varepsilon=\frac{U_0^2-U_c^2}{U_c^2}$ where $U_c$, the critical field, is given by the relation (\ref{eq:def-UC}) : 

\begin{equation}
U_c^2=\frac{\pi^2k_1}{\epsilon_a}
\label{eq:def-UC}
\end{equation}
The expression (\ref{eq:dvp_serie_mathematica}) of the free energy can be seen as the Ginzburg-Landau description of a second order phase transition with order parameter $\theta_m$ and control parameter $\varepsilon$. 
The minimization of this free energy as a function of $\theta_m$ gives the relation (\ref{eq:thetam2-eq}) between the mean value of the order parameter : $<\theta_m>$, ($<\cdot>$ stands for temporal averaging ) and the control parameter. 
\begin{equation}
<\theta_m>=\pm\sqrt{\frac{2\varepsilon}{\kappa+\varUpsilon+1}}
\label{eq:thetam2-eq}
\end{equation}

In this framework, one can also study the dynamics of the FT by introducing the rotational viscosity $\gamma$ and equating the frictional dissipation to the derivative of the free energy. Then, defining the characteristic time $\tau_0=\frac{\gamma L^2}{\pi^2 k_1}$, and adding thermal noise $\eta$, $\delta$-correlated in time, one gets the following Ginzburg-Landau equation\cite{Sanmiguel1985} to describe the dynamics of $\theta_m$ : 
\begin{equation}
\tau_0 \frac{{\rm d} \theta_m}{{\rm d} t}=  \varepsilon \ \theta_m -
{1 \over 2} (\kappa+\varUpsilon+1) \theta_m^3 +\eta
\label{momentum_equation}
\end{equation}
The equilibrium solution of (\ref{momentum_equation}) is  of course given by relation (\ref{eq:thetam2-eq}).

To study the dynamics of the fluctuations, we decompose the dynamics of $\theta_m$ in its mean value $<\theta_{m}>$ and its fluctuations $\delta\theta$ : $\theta_m(t)=<\theta_{m}>+\delta\theta(t)$, $<\theta_{m}>$ is known from (\ref{eq:thetam2-eq}). Introducing this decomposition in (\ref{momentum_equation}), we get the following equation for the fluctuations : 

\begin{equation}
\tau \delta\dot{\theta}=-2\varepsilon\delta\theta+\eta
\label{eq:delta-theta1}
\end{equation}

From this equation (that can be seen as a Langevin equation -- for more details, see \cite{joubaud-jstat,joubaud-gaussian}) one deduces the following expression for the variance  $\sigma^2_\theta$ of $\delta \theta$ above the critical point: 

\begin{equation}
\sigma_\theta^2=<\delta\theta^2>\propto \frac{k_BT}{2\varepsilon}
\label{eq:variance-theta}
\end{equation}
 
 Thus the amplitude of $\sigma_\theta^2$ diverges at critical point, as one would expect from a second order transition. This divergence gives us a good way to measure the critical point. 
However, to use this property, one has to take into account how the order parameter $\theta_m$ is measured in our experiment. In fact, we use a very sensitive interferometer  (see ref\cite{Bellon02} for details) depicted on fig\ref{fig1} : a polarized laser beam (wavelength $\lambda=632.8$nm) arriving on the liquid crystal cell is decomposed in the ordinary and extraordinary polarizations. These two polarizations gets a dephasing $\phi$ after the cell because of the optical anisotropy of the medium, which depends on how much the molecules have rotated. A Wollaston prism makes the two  polarizations interfere in the first detector, as depicted in  fig\ref{fig1}c). The electric signal out of the photodiodes is proportional to $\sin\phi$. In detector 2, a $\lambda/4$ plate is inserted before the Wollaston, which leads to an electric signal proportional to $\cos \phi$. Combining the signal of the two detectors, one can extract a signal proportional to $\phi$. 
This proportionality is the great advantage of our system on apparatus based on crossed polarizers which have been widely used to study this transition. 
In fact, this last method has the inconvenient of being sensitive to $\cos\phi$ and therefore its sensibility to the fluctuations depends on the mean value of $\phi$. Moreover, our method is also much more sensitive than capacitive methods in which the averaging area is too large to access the fluctuations.
  

 In our experiment, the measured observable is $\phi$ which is acquired with a resolution of $24$ bits at a sampling rate of $1024$ Hz. In the vicinity of the threshold, $\phi$ is proportional to the square of the order parameter $\theta_m$ \cite{joubaud-gaussian}. In fact, in this framework, $\phi$ and $\theta$ are related by the following expression : 
 \begin{equation}
\phi=\frac{\phi_0\nu}{4}\theta_m^2
\label{eq:phi-theta}
\end{equation}
 where $\phi_0=2\pi Ln_e/\lambda$, $\nu=\frac{n_e^2-n_o^2}{n_e^2}$ is the optical anisotropy with  $n_o$ and $n_e$   the refractive index of the ordinary and extraordinary polarizations ($\nu=0.2457$ in 5CB). 
To find the relation between the fluctuations of $\phi$ and those of $\theta_m$, we derivate equation (\ref{eq:phi-theta}) and obtain the relation  (\ref{eq:delta-phi-delta-theta}) between $\delta\theta$ and $\delta\phi$ which shows that, around the threshold, $\delta\phi$ is proportional to $<\theta_m>\delta\theta$ and not directly proportional to $\delta\theta$. 

\begin{equation}
\delta\phi=\frac{\phi_0\nu<\theta_{m}>}{2}\delta\theta
\label{eq:delta-phi-delta-theta}
\end{equation}

From eq(\ref{eq:delta-phi-delta-theta}), we built the variance $\sigma_\phi^2$, the one which is actually measured, as follows :  

\begin{equation}
\sigma_\phi^2=<\delta\phi^2>=\frac{\phi_0^2\nu^2}{4}<\theta_{m}>^2<\delta\theta^2>
\label{eq:var-phi}
\end{equation} 

In this expression, we see directly that before the threshold the variance of $\phi$ is zero. Above the threshold, $<\theta_{m}>^2$ is linear in $\varepsilon$ (cf eq(\ref{eq:thetam2-eq})) whereas $<\delta\theta^2>$ is proportional to $1/\varepsilon$ (cf eq (\ref{eq:variance-theta})), therefore $\sigma_\phi^2$ remains constant.
Taking into account the fact that $\phi$ is proportional to $\theta_m^2$ (see eq (\ref{eq:phi-theta})), $\sigma^2_\phi/<\phi>$ will have the same dependence in $\varepsilon$ as $\sigma^2_\theta$. 

\vspace{0.3cm}

\begin{figure}[htbp]%
\begin{center}
\includegraphics[width=.5\columnwidth]{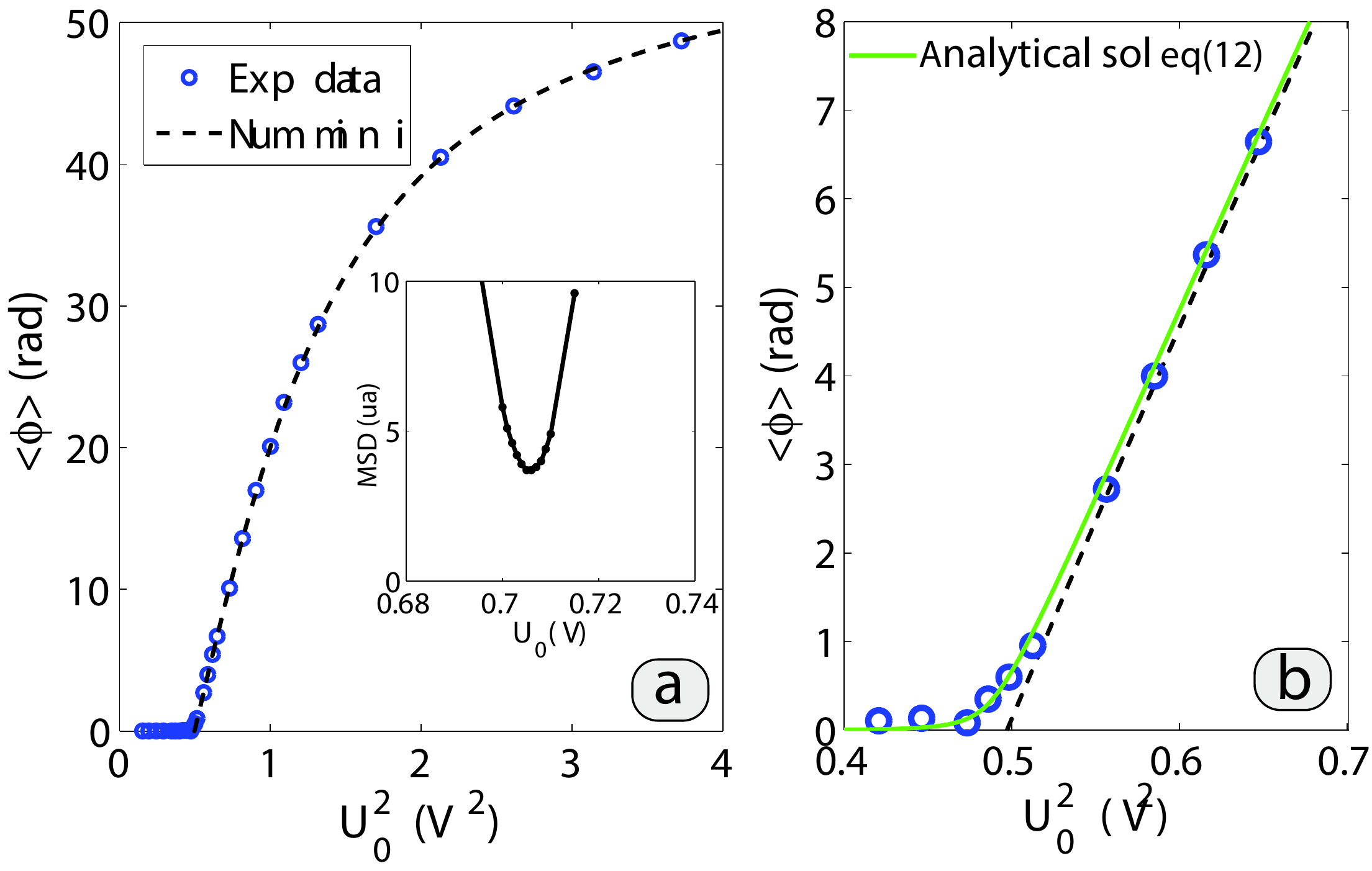}%
\end{center}
\caption{a) main figure : amplitude diagram of the FT. Inset : msd as a function of chosen $U_c$. b) Expansion around the threshold of the amplitude diagram}%
\label{fig2}%
\end{figure}

 Experimentally, we first measure $<~\phi~>$ as a function of the voltage $U_0$, applied to the cell, by averaging the acquired signal over a few minutes (the typical timescale of the system is about 10 seconds). We plot the experimental results on the main figure of fig\ref{fig2}a) and, as predicted, we observe an increase of the control parameter above a particular value of $U_0$, about 0.7V. 
 On the same figure, we also plot the result of the numerical integration of the minimization (Num-Min) of $F_s$. Let us recall that this calculation depends on one free parameter, $U_c$ (see ref.\cite{Deuling} for details), the thickness of the cell $L$ being entirely determined by the dephasing $\phi$ obtained at high voltage, when all molecules have rotated. The value of $U_c$ is therefore determined by computing the minimum square difference (msd) between the data and the fit obtained for a given value of $U_c$ \cite{Bradshaw, faetti-bogi}. The parabola in the inset of fig\ref{fig2}a) shows the values of the minimum square difference (msd) as a function of the chosen $U_c$. The minimum gives the best estimation of $U_c$, whereas the uncertainty is given by the curvature of the well (more precisely, $\Delta U^2=1/2C$ where C is the local curvature of the well). The resulting uncertainty  in our case is 0.003V.
\begin{equation}
U_c(\mbox{num})=0.705 \pm 0.003 V
\label{eq:Uc-num}
\end{equation}

An expansion around the critical point is shown on fig\ref{fig2}b) where it can be seen that the transition is not sharp as would be expected from a second order phase transition, instead, the data are characteristic of an imperfect transition : this is due to cell assembling and preparation. In fact, because the angle made with the director and the confining plates (pre-tilt angle) cannot be exactly zero, we choose an anti-parallel configuration (see fig\ref{fig1}a-b or ref\cite{yeh-book} ; direction of rubbing of both plates is the same) where, in the ideal case, the equilibrium configuration in the absence of external field is symmetric. Nevertheless, a truly symmetric assembling of the cell is not possible, and there is always a small residual angle between the plates that facilitates the transition. 
To explain the roundness of the transition, the small asymmetry between the plates can be taken into account phenomenologically by adding a constant term to eq(\ref{momentum_equation}) : 
\begin{equation}
\tau\dot{\theta}_m=\varepsilon\theta_m-{1 \over 2} (\kappa+\varUpsilon+1) \left(\theta_m^3-\theta_0^3\right) +\eta
\label{eq:dyn_theta0}
\end{equation}
Experimentally, one gets $\theta_0^3=0.0019$ rad$^3$ from the value of $\phi$ at the threshold. An implicit solution of eq(\ref{eq:dyn_theta0}) is calculated and plotted on fig\ref{fig2}b) (green continuous line). We see that the roundness of the transition is well explained by eq(\ref{eq:dyn_theta0}).
This unavoidable asymmetry of about ten percent on $\theta_s$ will also be responsible for a cut-off in the divergence of $\sigma_\theta^2$ and a smoothing of the $\sigma^2_\phi$ curve as it is developed in ref\cite{Caussarieu1}.

\begin{figure}[htbp]
\includegraphics[width=0.5\columnwidth]{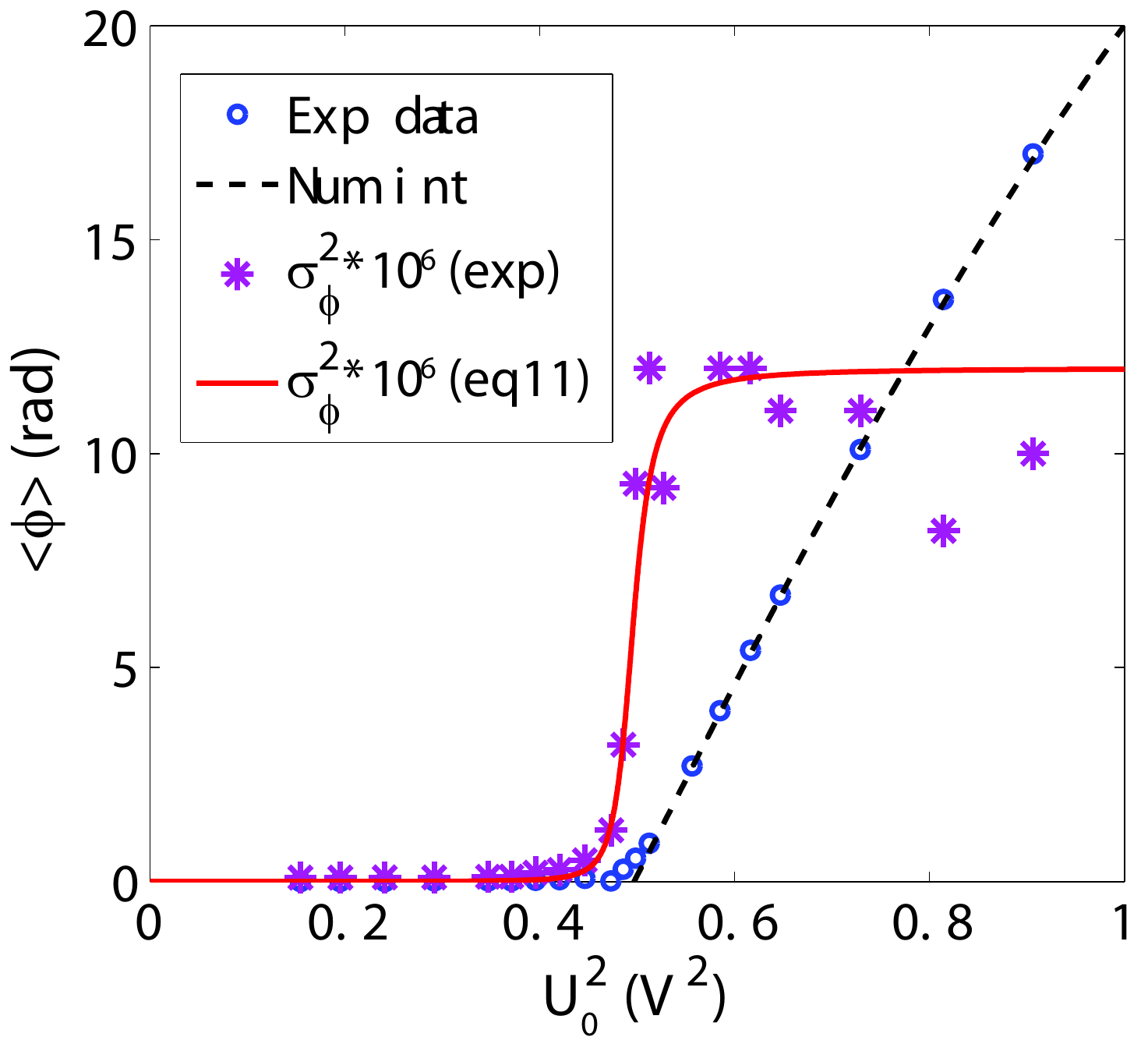}%
 \caption{Phase transition diagram and evolution of the variance of $\phi$ as a function of $U_0$}%
 \label{fig3}
 \end{figure}

In our system, the typical amplitude of the fluctuations is 3 order of magnitude smaller than the mean value of the acquired signal. For this reason, it is impossible to avoid any drift over few hours of acquisition, and the estimation of the variance of the static signal ($\sigma^2_\phi=<(\delta\phi)^2>$) cannot be done over a single acquisition. The best compromise we found is to estimate the variance on a temporal average over different acquisitions of about 3 minutes in which the slow drift cannot be perceived, and then, averaging in a statistical way on the independent measurements of the variance. We report $\sigma^2_\phi$ as a function of $U_0^2$ on fig\ref{fig3}. 

%

On the same figure, we also plot (red continuous line) the variance obtained from eq(\ref{eq:dyn_theta0}), and a good agreement is observed with the experimental data. The decrease observed for large $\varepsilon$ is due to the non linearities as discussed in \cite{Caussarieu1}. Such a behavior has already been observed in ref\cite{Zimmermann,Zhou2004} using shadowgraph methods, \alb{and in ref\cite{Galatola1992, Galatola1994} with light diffusion methods,} but has never been explained (in ref\cite{Zhou2004}, a first order transition has been proposed to explain the continuous growth of the fluctuations at the threshold even if no hysteresis was found). 

\begin{figure}[h!]
\includegraphics[width=0.5\columnwidth]{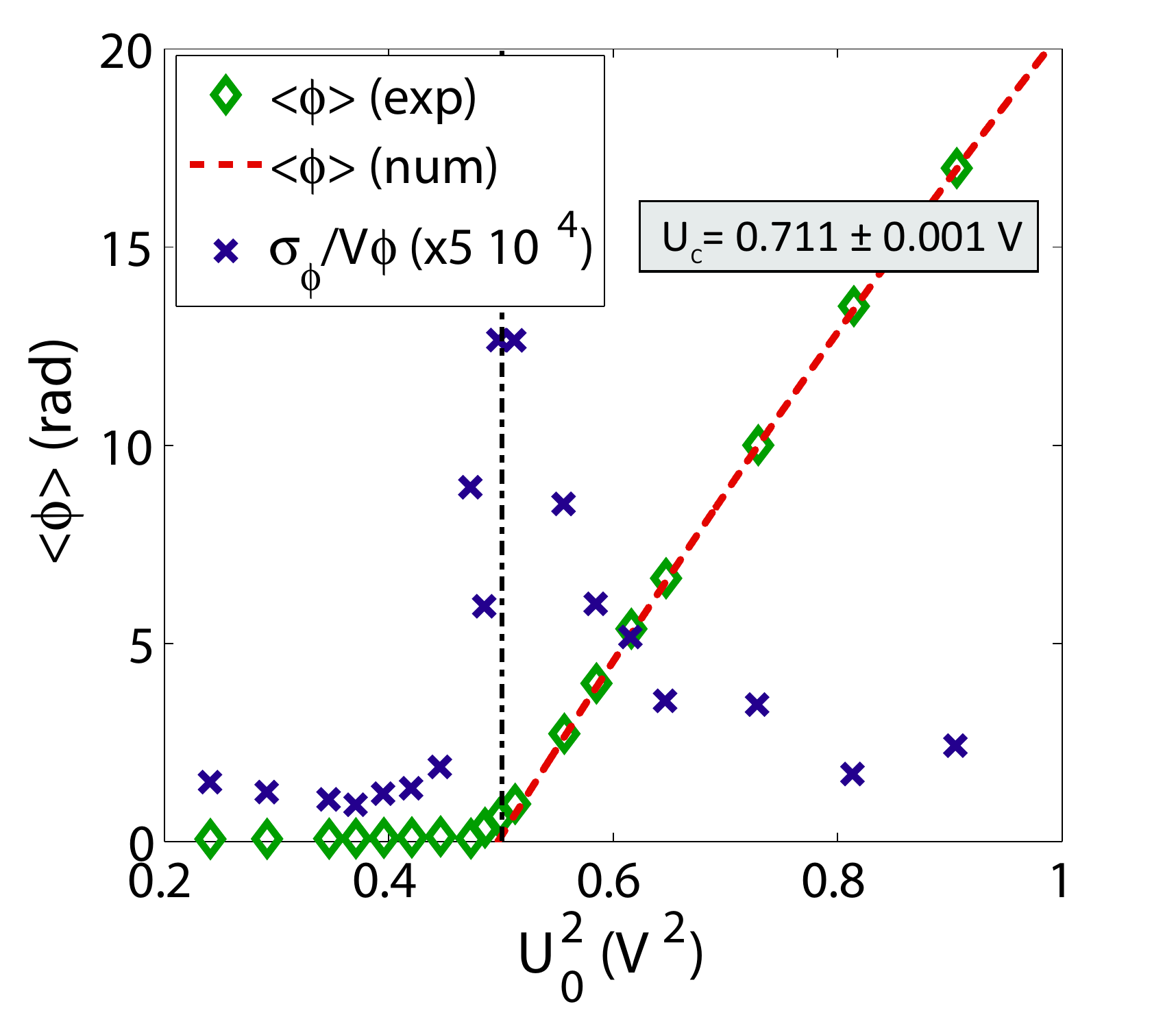}%
 \caption{Phase transition diagram and evolution of the variance of $\phi$ divided by the square root of the mean value of the dephasing $\phi$}%
 \label{fig4}
 \end{figure}

On figure \ref{fig4}, we plot $\sigma^2_\phi/\phi$ which, thanks to eq(\ref{eq:var-phi}), shall diverge for $\varepsilon=0$, that is to say, for $U_0=U_c$.


Indeed, we can clearly see that there is a sharp maximum which stands much more closely to the apparent beginning of the instability. This maximum gives us a new definition of the critical threshold : 
 $$U_c(fluct)=0.711\pm 0.001$$

  The difference between the two measurement of the electrical threshold is due to the fact that, in our numerical integration, we neither take into account the effect of the finite anchoring energy nor the existence of a small asymmetry in the cell which both tend to lower the threshold. When comparing our results to those of Faetti \cite{faetti-bogi}, taking into account our distance to the nematic-isotrop transition ($\Delta T = 10$ Celsius Degree), we get $U_c=0.710$ which is very close to our measurement based upon fluctuations. 

 To conclude, we have described a new method for measuring the electrical threshold of the Fr\'eedericksz transition. {The great advantage of the method is that it does not require any complex fit and any hypothesis on boundary conditions. Furthermore  we have, for the first time, a solid explanation for the constant growing of the fluctuations while passing the threshold measured with different optical sets up. 
 \vskip 30pt
 This experiment is supported by the ERC grant OUTEFLUCOP. We acknowledge useful discussion with P. Oswald }


\begin{thebibliography}{0}%
\makeatletter
\providecommand \@ifxundefined [1]{%
 \@ifx{#1\undefined}
}%
\providecommand \@ifnum [1]{%
 \ifnum #1\expandafter \@firstoftwo
 \else \expandafter \@secondoftwo
 \fi
}%
\providecommand \@ifx [1]{%
 \ifx #1\expandafter \@firstoftwo
 \else \expandafter \@secondoftwo
 \fi
}%
\providecommand \natexlab [1]{#1}%
\providecommand \enquote  [1]{``#1''}%
\providecommand \bibnamefont  [1]{#1}%
\providecommand \bibfnamefont [1]{#1}%
\providecommand \citenamefont [1]{#1}%
\providecommand \href@noop [0]{\@secondoftwo}%
\providecommand \href [0]{\begingroup \@sanitize@url \@href}%
\providecommand \@href[1]{\@@startlink{#1}\@@href}%
\providecommand \@@href[1]{\endgroup#1\@@endlink}%
\providecommand \@sanitize@url [0]{\catcode `\\12\catcode `\$12\catcode
  `\&12\catcode `\#12\catcode `\^12\catcode `\_12\catcode `\%12\relax}%
\providecommand \@@startlink[1]{}%
\providecommand \@@endlink[0]{}%
\providecommand \url  [0]{\begingroup\@sanitize@url \@url }%
\providecommand \@url [1]{\endgroup\@href {#1}{\urlprefix }}%
\providecommand \urlprefix  [0]{URL }%
\providecommand \Eprint [0]{\href }%
\providecommand \doibase [0]{http://dx.doi.org/}%
\providecommand \selectlanguage [0]{\@gobble}%
\providecommand \bibinfo  [0]{\@secondoftwo}%
\providecommand \bibfield  [0]{\@secondoftwo}%
\providecommand \translation [1]{[#1]}%
\providecommand \BibitemOpen [0]{}%
\providecommand \bibitemStop [0]{}%
\providecommand \bibitemNoStop [0]{.\EOS\space}%
\providecommand \EOS [0]{\spacefactor3000\relax}%
\providecommand \BibitemShut  [1]{\csname bibitem#1\endcsname}%
\let\auto@bib@innerbib\@empty
\end{thebibliography}%


\begin{thebibliography}{0}

\bibitem{faetti-bogi} A. Bogi S. Faetti, Liquid Crystals 28, 729 (2001). 

\bibitem{Bradshaw} M.J. Bradshaw, E.P. Raynes, J.D. bunning and T.E. Faber, {\em J. Physique} {\bf 46}, (1985) 1513-1520

\bibitem{DeGennes}P.G. de Gennes and J. Prost, {\em The physics of liquid crystals}
(Clarendon Press, Oxford, 1974).

\bibitem{Oswald}
P. Oswald and P. Pieranski, {\em Nematic and cholesteric liquid
crystals} (Taylor \& Francis, 2005).

\bibitem{cognard} 
J. Cognard, \textsl{Alignment of nematic liquid crystals and their mixtures} (Gordon and Breach science publishers, 1982)

\bibitem{yeh-book} 
P. Yeh and C. Gu, \textsl{Optics of liquid crystal displays} (John Wiley $\&$ Sons, Inc , 1999)

\bibitem{franck} F. Franck, Discuss. Faraday soc. , \textbf{25}, 19 (1958)

\bibitem{Deuling} H.J. Deuling,  Molecular crystals and liquid crystals, 19, 123 (1972).

\bibitem{Sanmiguel1985}
M. San Miguel, {\em Phys. Rev. A} {\bf 32}(6) 3811 (1985).

\bibitem{joubaud-jstat}
S. Joubaud, A. Petrosyan, S. Ciliberto, {\em J. Stat. Mech.}, P01033(2009) 

\bibitem{Bellon02} L.\ Bellon, S.\ Ciliberto, H.\ Boubaker, L.\ Guyon, {\em Optics Communications} {\bf 207} 49-56 (2002).

\bibitem{joubaud-gaussian}
S. Joubaud, A. Petrosyan, S. Ciliberto, N. Garnier,  {\em Phys. Rev. Let.} {\bf 100}(18) 180601 (2008).

\bibitem{Caussarieu1} 
A. Caussarieu, A. Petrosyan, S. Ciliberto, Physical Review letters submitted arxiv  

\bibitem{Zimmermann} 
B. L. Winkler, H. Richter, I. Rehberg, W. Zimmermann, L. Kramer, and A. Buka. {\em Phys. Rev. A}, {\bf 43}, 4 (1991)

\bibitem{Zhou2004}
S.-Q. Zhou and G. Ahlers, arXiv:nlin/0409015v2 (2004)

\bibitem{Galatola1992}
P. Galatola, {\em J. Phys. II France} {\bf 2} 1995 (1992).

\bibitem{Galatola1994}
P. Galatola and M. Rajteri, {\em Phys. Rev. E} {\bf 49} 623
(1994).

\bibitem{guyon} P. Pieranski, F. Brochard, and E.Guyon,  {\em J.physique},
33, 68, 1972;   P. Pieranski, F. Brochard, and E.Guyon, {\em J. Phys. II, France} 34, 35, 1973.

%
%
%
%






%






\end{thebibliography}

\end{document}